\documentclass[preprint,aps,groupedaddress,amsmath,amssymb]{revtex4-1}
\usepackage{bm}% bold math
\usepackage{color,graphicx}
\usepackage{subfig}
\begin{document}

\title{Classical understanding of the electron vortex beams in a uniform magnetic field}

\author{Yeong Deok Han}
\affiliation{Department of Computer Science and Engineering, Woosuk University, Wanju, Cheonbuk, 565-701, Republic of Korea}
\author{Taeseung Choi}
\email{tschoi@swu.ac.kr}
\affiliation{ Division of Applied Food System, College of Natural Science, Seoul Women's University, Seoul 139-774, Republic of Korea} %\textbackslash\textbackslash
\affiliation{School of Computational Sciences, Korea Institute for Advanced Study, Seoul 130-012, Korea}

%
%\date{\today}

\begin{abstract}

Recently interesting observations on electron vortex beams, which have angular momentum about the center of the vortex beams, 
have been made. 
We have shown that the basic features of the electron vortex beams in a uniform magnetic field are understandable by using 
the classical motions of electrons. 
We have constructed a classical vortex-like motion by the collective motion of 
individual electrons in their cyclotron motions with a constant canonical angular momentum in the symmetric gauge, which models 
electron vortex beams, in a uniform magnetic field. 
With this model the various properties of circulating currents and the relation between energy and kinetic angular momentum 
in the electron vortex beams are well explained. 
We have also shown that the mismatch between the centers of the electron vortex beam and the classical cyclotron orbits naturally induces 
the parallel axis theorem and also the time-varying kinetic angular momentum of the electron vortex beam for certain 
distributions of classical electrons. 

%This explains the reason why the electron vortex states are solved in the 
%symmetric gauge, not in the Landau gauge., even though the physics should be gauge invariant.  

\end{abstract}

%\pacs{ 03.65.-w, 03.65.Ta, 03.67.-a}% PACS, the Physics and Astronomy
\maketitle

Free electron vortex states have recently been predicted by considering semiclassical (paraxial) 
wave packet \cite{Bliokh07} and observed in electron microscopy \cite{Uchida, McMorrain, Verbeeck10}. 
The free electron vortex is practically equivalent to the optical vortex beam, 
however, in the presence of a magnetic field, the properties of electron vortex beam 
becomes different from those of its optical counterpart. 
The electron vortex beam with angular momentum has a magnetic moment and interacts with an external magnetic field, 
which gives interesting physics and applications \cite{Bliokh07, Verbeeck11, VerbeeckUl, Gallatin12, SchttschneiderNC, 
Guzzinati, Bliokh11, Lloyd, Karimi}. 

For the electron vortex beam, the interactions between electrons are assumed to be so small that can be neglected. 
Therefore, the problem of electron vortex beams in a magnetic field seems to be reduced to the problem of one electron 
in the same magnetic field. 
The physics of an electron in a uniform magnetic field is well understood both classically and quantum mechanically 
\cite{Ballentine}.  
The motion of a classical electron in a uniform magnetic field is circular with constant speed, known as the cyclotron 
motion, hence it is expected that the kinetic angular momentum of such motion about its center is constant.  
However, one of interesting issues, which seems to be contrary to classical cyclotron motion, was pointed out by 
Greenshields et al. \cite{Greenshields}. They showed that the "diamagnetic" angular momentum of the electron vortex beam 
in a uniform magnetic field is time-varying in general, which implies that 
the kinetic angular momentum of the electron vortex beam could also be time-dependent. 
They used the quantum vortex solutions to show the time-dependence of the kinetic angular momentum. 
This fact seems to be contradictory to the rotational symmetry of the system and they showed that the conservation of 
kinetic angular momentum is recovered by involving the angular momentum of electromagnetic field. 
%Their resolution reconciled with the rotational symmetry of the system.  
%And they used the vortex solution to show the time-dependence of quantum kinetic angular momentum.  
%Hence, it is not clear that the time dependence of kinetic angular momentum of an electron in a uniform magnetic field 
%is originated from the pure quantum nature of one electron or from the vortex property. 

It is, however, still surprising how the rotationally 
symmetric vortex solutions can have time-varying radius in quantum vortex solutions, contrast to the fact that 
the radius of the rotationally symmetric classical cyclotron motion is constant. 
It was found that the average radial position of the electron vortex state expands and 
contracts \cite{Gallatin12}. % and this contrasts with the fact that the radius of the classical cyclotron orbit is constant. 
%This indicates that the time-varying angular momentum induced by the time dependence of radial position 
%probably originates from the properties of an electron vortex beam. 
Recently it was also shown that the orbital angular momentum of an electron vortex beam can 
be decomposed into separate angular momenta according to parallel axis theorem, which seems to be only 
meaningful for an extended probability distribution \cite{Greenshields15}. 
%However, an electron vortex beam with torus-like intensity profile 
%cannot be created by the classical cyclotron motion of one electron, especially when the center of 
%the cyclotron orbit does not coincide with the center of the vortex beam. 

As we have seen above, the characteristics of the electron vortex beams are different from those of the classical 
cyclotron motion of one electron. 
%As far as we know, there is no classical model of an electron vortex beam. 
In this paper, we will show that it is possible to explain the motion of the electron vortex beams as a 
collective motion of electrons in their classical cyclotron motions. 
%construct the classical wave model for an electron vortex beam in a uniform magnetic field. 
Using this picture, the basic features of the electron vortex beams are well explained. 
We will also show that the time-varying behavior of the kinetic angular momentum and the parallel axis theorem 
can be understood.  
%as common properties of both an electron and an electron vortex beam, which originate 
%from the mismatch between the center of the classical cyclotron motion and the origin of the coordinate. 
%And we will also show that the electron vortex beams can be modeled by the classical waves constructed 
%by electrons in their individual cyclotron motion. 
%The time dependent property of kinetic angular momentum is also retained in the classical electron vortex beams. 
This suggests that our classical picture is very instructive to explain the electron vortex beams 
intuitively.   
%The three circular frequencies, Larmor, double Larmor, and zero of the Bohmian path of electron vortex states 
%can be also understood by the classical waves generated by the composition of circular motions about the 
%origin of the coordinate.    

\section{One electron motion in a uniform magnetic field}

The motion of one electron in a uniform magnetic field is a textbook problem, which is well understood 
both classically and quantum mechanically \cite{Ballentine}. 
However, not much attention is paid to the time-dependence of angular momenta, especially in classical point 
of view. And the classical motion of one electron will be the building block of the collective motion 
to explain the electron vortex beams. 
Hence, we will briefly review the problem of the motion of one electron in a uniform magnetic field as we focus on 
the time dependence of angular momenta.  

The model Lagrangian for an electron in a uniform magnetic field ${\bf B}=B \hat{\bf z}$, 
in a cylindrical coordinate ($\rho$, $\phi$, $z$) 
and the symmetric gauge with ${\bf A}_s=(B \rho /2 )\hat{\boldsymbol{\phi}}$ is as follows 
\begin{eqnarray}
\mathcal{L}&=&\frac{1}{2}m {\bf v}^2 + e {\bf v}\cdot {\bf A}_s \\ \nonumber
&=& \frac{1}{2}m \dot{\rho}^2 + \frac{1}{2}m \rho^2 \dot{\phi}^2 + \frac{1}{2} m \dot{z}^2+ \frac{e}{2}B \rho^2 \dot{\phi},
\end{eqnarray}
where $e$, $m$, and ${\bf v}$ are the charge, the mass, and the velocity of the electron, respectively. 
Then the canonical conjugate momenta are defined as
\begin{eqnarray}
p_\rho =\frac{\partial \mathcal{L}}{\partial \dot{\rho}}= m \dot{\rho}, ~~~
p_\phi=\frac{ \partial \mathcal{L}}{\partial \dot{\phi}}= m\rho^2 \dot{\phi}+ \frac{e}{2}B\rho^2.
\end{eqnarray}
This Lagrangian has a rotational symmetry about $z$-axis, i.e., there is no $\phi$-dependence, so that 
the canonical conjugate momentum $p_\phi$ is constant of motion. 
Note that $p_\phi$ is the $z$-component of the canonical angular momentum, i.e., 
$p_\phi\equiv {L}_z=({\bf r}\times {\bf p})_z$. 
That is, the canonical angular momentum $L_z$ of the electron in a uniform magnetic field is conserved 
in the symmetric gauge.  
The $L_z$, however, is gauge dependent. 

The gauge invariant kinetic angular momentum is defined as 
%gives the energy of the interaction with the uniform magnetic field. 
\begin{eqnarray}
\label{eq:KAM}
L_z^{\mbox{\scriptsize{kin}}}= {\bf r}\times ({\bf p}-e{\bf A}_s)=L_z + \frac{m \omega_c}{2}\rho^2, 
\end{eqnarray}
where ${\bf r}=\rho \hat{\rho}+z \hat{z}$ and $\omega_c=-eB/m$ is the classical cyclotron frequency of the electron. 
The second term in Eq. (\ref{eq:KAM}) is called the diamagnetic angular momentum.
%Hence the conservation of the kinetic angular momentum depends on the time-dependence of 
%the diamagnetic angular momentum, which is determined by 
The conservation of the canonical angular momentum is guaranteed by the rotational symmetry of the Lagrangian, 
on the other hand, a kinetic angular momentum depends on the origin of the coordinate.  
This means that the conservation of a kinetic angular momentum in one coordinate system does not automatically 
guarantee the conservation of the kinetic angular momentum in another coordinate system with a different 
origin. 

To study the effect of the choice of the origin of the coordinate, 
%That is, even the cyclotron motion has been considered to have constant kinetic angular momentum classically 
%in the coordinate, where the center of the cyclotron motion is the origin of the coordinate, the kinetic angular 
%momentum can be time dependent in another coordinate with a different origin. 
the Cartesian coordinate is convenient.  
The Hamiltonian $\mathcal{H}={\bf v}\cdot {\bf p}-\mathcal{L}$ is written in the Cartesian coordinate as
\begin{eqnarray}
\label{eq:HCT}
\mathcal{H}=\frac{1}{2m} {\bf p}^2 +\frac{\omega_c}{2} L_z + \frac{m}{8}\omega_c^2 (x^2+y^2),
\end{eqnarray}
where ${\bf p}^2=p_x^2+p_y^2+p_z^2$. 
The classical equation of motion for the electron becomes 
the usual Lorentz force equation, 
\begin{eqnarray}
\label{eq:LFE}
m\frac{d^2}{dt^2} {\bf r}= e {\bf v}\times {\bf B}.
\end{eqnarray}
The $z$-directional classical motion of the electron described by Eq. (\ref{eq:LFE}) is trivial 
because there is a translation symmetry along $z$-direction. 
Hence we focus on the classical motion of the electron in the $xy$-plane. 

Eq. (\ref{eq:LFE}) describes the cyclotron motion in the $xy$-plane and the general solutions are
\begin{eqnarray}
\label{eq:2DSOL}
x(t)= x_0 + R \cos{(\omega_c t+\theta)}, ~~~y(t)=y_0+R\sin{(\omega_c t+\theta)},
\end{eqnarray}
where $(x_0, y_0)$ is the center of the cyclotron orbit with radius $R$ and 
$\theta$ is the phase shift from the $+x$ direction. 
Then the squared 2-dimensional radial distance $\rho^2=x^2+y^2$ of the electron 
from the origin of the coordinate becomes 
\begin{eqnarray}
\label{eq:SRD}
\rho^2&=&x_0^2+y_0^2+R^2 \\ \nonumber
&+&2x_0  R \cos{(\omega_c t+\theta)}+ 2y_0 R\sin{(\omega_c t+\theta)},
\end{eqnarray}
which shows time dependence.  
By direct calculation one obtains the 
following dynamical equation for the squared 2-dimensional radial distance as
\begin{eqnarray}
\label{eq:CRDE}
\frac{d^2}{dt^2}\rho^2=  - \omega_c^2 \rho^2 - 2 \frac{\omega_c}{m} L_z +\frac{4}{m}E, 
\end{eqnarray}
where $E$ is the 2-dimensional energy (classical Hamiltonian) $\frac{1}{2}m(v_x^2+v_y^2)=\frac{1}{2}mR^2\omega_c^2$. 
One can easily check that Eq. (\ref{eq:CRDE}) is equivalent to quantum mechanical equation Eq. (11) 
for the squared radius in Heisenberg formalism of Ref. \cite{Greenshields}.  

This suggests that the oscillations of the 2-dimensional squared radial distance and the resultant time-varying 
diamagnetic response could be understood as originated from the mismatch of the origin of the coordinate and the center 
of the cyclotron orbit. 
% not real physical effect independent of the choice of the origin of 
%the coordinate, but originate just from the mismatch of the origin of the coordinate and the center 
One can also check that the classical torque ${\bf r}\times (e{\bf v}\times {\bf B})$ is not zero 
for the cyclotron motion of the electron with centers different from the origin of the coordinate.  
%because the Lorentz force $e{\bf v}\times {\bf B}$ is always directed to the center of the cyclotron orbit 
%not to the origin of the coordinate. 
This torque is responsible for the change of the kinetic angular momentum of the electron. 
If the origin of the coordinate and the center of the cyclotron orbit match, i.e., 
$x_0=y_0=0$, $\rho^2$ ($=R^2$) is 
a constant of motion and the kinetic angular momentum is also conserved as well as the canonical 
angular momentum. 
%When the uniform magnetic field is in the entire $xy$ plane, 
%every points in the $xy$ plane are indistinguishable such that the most natural choice of the origin 
%of the coordinate is the center of the cyclotron orbit. Then the rotational symmetry of the physical motion 
%is manifest and the kinetic angular momentum will be conserved by itself without considering the angular momentum of 
%the electromagnetic field.   
The time dependence of the kinetic angular momentum of the classical electron vortex beam will be discussed 
in sec. \ref{sec:CEVB}. 

%\textcolor{red}{ {\bf Must be included??}
%When the uniform magnetic field exists in the finite circular region with definite center, then 
%the position of the center of circular orbit of the electron affects the kinetic angular 
%momentum physically such that the kinetic angular momentum oscillates in time and is not 
%conserved. Howevre, in this case we can obtain the conserved total angular momentum consisting 
%of the kinetic angular momentum of the electron and the field angular momentum.  }

\section{Classical model of an electron vortex beam in a uniform magnetic field} 
\label{sec:CEVB}

In this section we will study the collective motions of classical electrons to show 
the basic features of the electron vortex beams. 
We will call the collective motions showing vortex-like motion a classical electron vortex.

%constructed by the electrons moving in their individual cyclotron orbits. 
% The time-varying kinetic angular momentum has been explained in both symmetric and Landau gauges 
%for one electron, on the other hand, 
%The physics should be invariant under the choice of different gauges, hence it is curious about 
%why the electron vortex beam physics can be understood only by using the symmetric gauge.  
 
%which is constructed 
%by a collective motion of each water molecule in its circular motion. %, not the de Broglie matter wave of an electron. 

We consider that electrons are moving with the same speed in a uniform magnetic field and 
there is no Coulomb repulsion between electrons, which is assumed in the usual electron vortex beams. 
Then %initial speeds of all electrons are assumed to be equal, hence, 
all electrons rotate in their cyclotron orbits with the same radius. 
The cyclotron orbits with the same radius $R$ given by the solutions in Eq. (\ref{eq:2DSOL}) 
can be classified by the canonical angular momentum in the symmetric gauge, because the canonical angular momentum 
is constant of motion.

\begin{figure}[htbp]
\subfloat[]{
\includegraphics[height=1.0in]{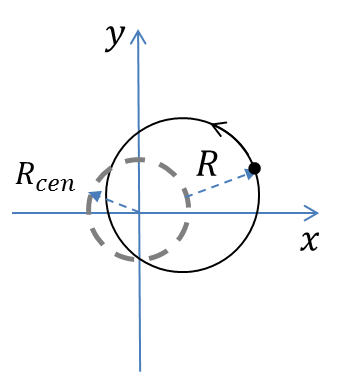}}
%\caption{The cyclotron orbit for $L_z<0$}
\subfloat[]{
\includegraphics[height=1.0in]{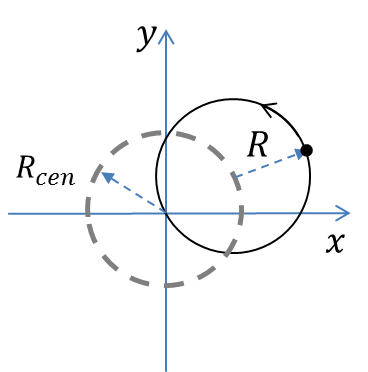}}
\subfloat[]{
\includegraphics[height=1.0in]{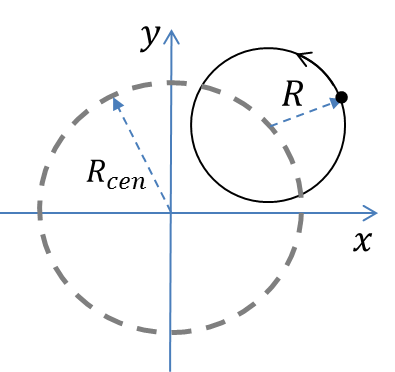}}
\caption{The cyclotron orbits for (a) $L_z>0$, (b) $L_z=0$, and (c) $L_z<0$.}
\label{fig:CAM}
\end{figure}

The canonical angular momentum $L_z$ of the electron on its cyclotron orbit, the $z$-component of 
${\bf r} \times (m {\bf v}+ e {\bf A}_s)$, is calculated as
\begin{eqnarray}
\label{eq:CCAM}
L_z= \frac{m}{2}\omega_c (R^2 -R_{\mbox{\scriptsize{cen}}}^2),
\end{eqnarray} 
where $R_{\mbox{\scriptsize{cen}}}$ is the distance from the origin of the coordinate to the center of 
the cyclotron orbit, i.e., $R_{\mbox{\scriptsize{cen}}}^2=x_0^2+y_0^2$. % and $R$ is the cyclotron radius. 
Hence there are three categories in the cyclotron motions of the electrons according to $L_z>0$, $L_z=0$, 
and $L_z<0$ as in Fig. \ref{fig:CAM}. 
The canonical angular momentum $L_z$ for the cyclotron motion of the electrons with the same radius $R$ 
becomes positive for $R^2>R_{\mbox{\scriptsize{cen}}}^2$, 
zero for $R^2=R_{\mbox{\scriptsize{cen}}}^2$, and negative for $R^2< R_{\mbox{\scriptsize{cen}}}^2$.  
%This implies that the cyclotron motions with the same cyclotron radius can be classified by 
%the sign of the canonical angular momentum depending on the distance $R_{\mbox{\scriptsize{cen}}}$, 
%which is also time-independent. 
%Note that the $L_z$ in the symmetric gauge is constant of motion, and $R$ and $R_{\mbox{\scriptsize{cen}}}$ is 
%not time-dependent. 

\begin{figure}[htbp]
\subfloat[]{
\includegraphics[height=1.0in]{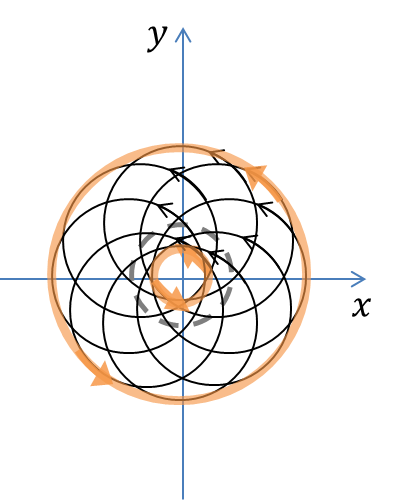}}
%\caption{The cyclotron orbit for $L_z<0$}
\subfloat[]{
\includegraphics[height=1.0in]{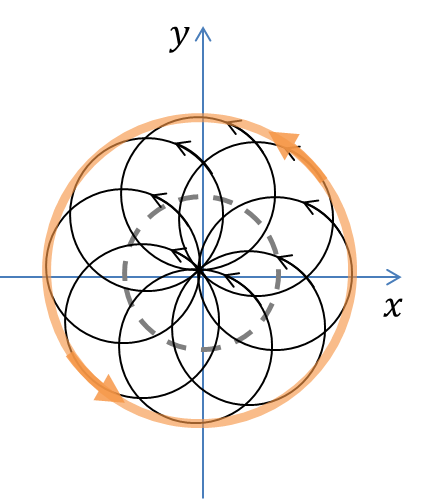}}
\subfloat[]{
\includegraphics[height=1.0in]{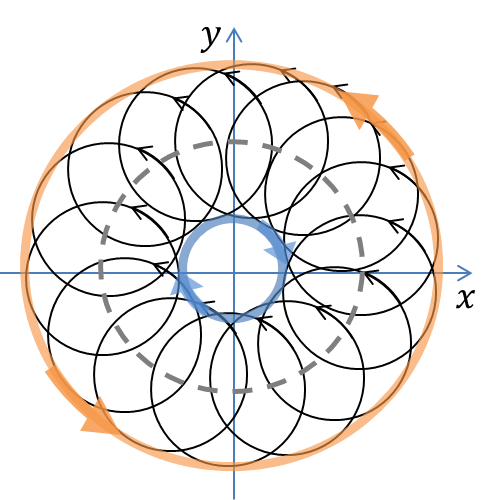}}

\subfloat[]{
\includegraphics[height=1.0in]{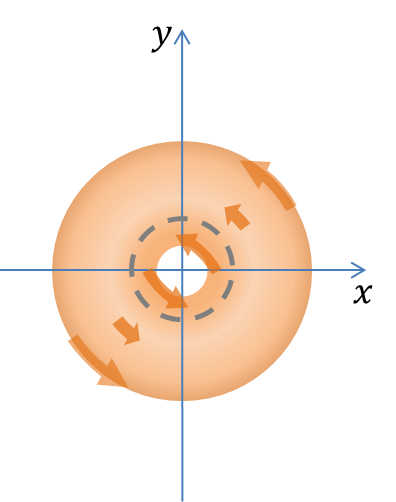}}
%\caption{The cyclotron orbit for $L_z<0$}
\subfloat[]{
\includegraphics[height=1.0in]{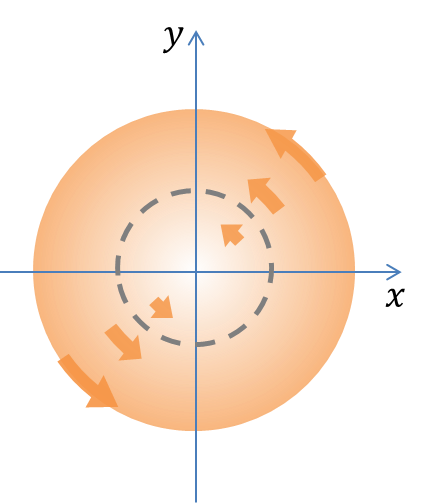}}
\subfloat[]{
\includegraphics[height=1.0in]{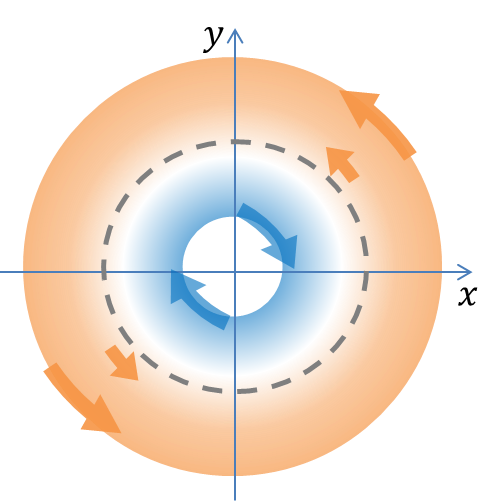}}
\caption{(Color online) The construction of the classical electron vortices by the collective motions of 
the constituent electrons in their own cyclotron orbits for 
(a) $L_z>0$, (b) $L_z=0$, and (c) $L_z<0$ and the resultant classical electron vortices with torus-like profile for 
(d) $L_z>0$, (e) $L_z=0$, and (f) $L_z<0$. The azimuthal frequencies of the circulating currents are 
$\omega_c$ ((a) and (d)), $\omega_c/2$ ((b) and (e)), and zero ((c) and (f)), respectively.  }
\label{fig:CEV}
\end{figure}

Here we construct the classical electron vortex by the collective motion of the electrons with the same canonical 
angular momentum $L_z$.  
%The assumption that the initial transverse motion of the uniformly distributed electrons are the 
%same is reasonable. Then the radius $R$ of all cyclotron orbits are the same, so that 
%the distances $R_{\mbox{\scriptsize{cen}}}$ of the all cyclotron orbits for the same $L_z$ are the same. 
The centers of the cyclotron orbits with the same $L_z$ have the same distance to 
the origin of the coordinate. 
We assume that the distribution of the cyclotron orbits are rotationally symmetric about the origin of the coordinate 
and the electrons are also uniformly distributed on these cyclotron orbits as shown in Fig \ref{fig:CEV} (a), (b), and (c).

As a result, the average current of the collective motions of the electrons is created, 
which is the rotational current around the center of the coordinate as shown in Fig \ref{fig:CEV} (d), (e), and (f).  
This apparent motion is the vortex-like motion with 
torus-like profile, which is a rotation around the center of the vortex as an average motion. 
However, the actual motion of individual electrons constructing the classical electron vortex is not rotation 
about the vortex center, but the cyclotron motion about its own center.

The classical electron vortex can be considered as a classical wave, in the similar sense to a water wave, 
in which the apparent motion of water wave is translational (passing to a certain direction), 
but each water molecule creating the water wave executes its own circular motion. 

There are three kinds of collective motions as shown in Fig. \ref{fig:CEV} 
according to the three categories of the $L_z$ in Fig. \ref{fig:CAM}. 
The collective motion of constituent electrons generates azimuthal current by averaging their 
individual motions. 
For $L_z>0$, all azimuthal currents are counter-clockwise and the magnitudes of the inmost and the outmost currents are 
given by the same and the maximum average speed. 
For $L_z=0$, the magnitude of the azimuthal currents reduces from the maximum value of the outmost current to zero 
at the origin of the coordinate, which is the center of the classical electron vortex. 
For $L_z<0$, the azimuthal current changes its sign at the point between the inmost and the outmost edges. 
Therefore, the azimuthal currents on the outer side and the inner sides are counter-clockwise and clockwise, respectively. 
The three different types of azimuthal currents are qualitatively 
equivalent to the types of the azimuthal currents in Fig. 5 of Ref. \cite{BliokhX}. 
This shows that the basic rotating features of the electron vortex beams can be understood by the classical 
electron vortex created by individual cyclotron motions.    

One of the intriguing result is that the the azimuthal angular frequencies of the electron vortex beams are 
cyclotron, Larmor, and zero frequencies according to the three categories of the canonical angular 
momentum \cite{SchttschneiderNC}. 
The expectation value of azimuthal angular frequency of the electron vortex beam is determined by 
$\langle \omega \rangle= \langle j /r \rangle$, where 
${\bf j}= \hbar/m [ \mbox{Im}(\psi^* {\bf j} \psi ) - e {\bf A}_s |\psi|^2 ]$ and $\psi$ is the vortex solution. 
Hence, the azimuthal angular frequency is related with the circulating current around the vortex center by 
$I_c= \langle \omega \rangle/(2\pi)$.
The three circulating currents given by the three azimuthal angular frequencies can also be explained by 
the classical electron vortex.  

%The circulating currents of electron vortex beams should be considered by the circulating current of 
%each individual cyclotron motion, which constructs the classical electron vortex, because the motions of electrons 
%constructing the classical electron vortex do not actually move along the apparent motion of the vortex, but along 
%their own cyclotron orbits. 
%Then the circulating currents of the electron vortex beams depend on the sign of the canonical angular momentum. 
The frequency of the azimuthal current of the classical electron vortex is determined by the angle 
through which each constituent electron rotates around the center of the vortex.  
The cyclotron orbits of electrons with $L_z>0$ encloses the origin of the coordinate. 
This implies that the electron circles around the origin of the coordinate every time the electron circles 
along its own cyclotron orbit. 
Hence the frequency of the circulating current of the electron vortex beam becomes the cyclotron 
frequency $\omega_c$. 
On the other hand, the cyclotron orbits of electrons with $L_z<0$ does not enclose the origin of the coordinate. 
This means that the circulating current of the electron vortex beam is zero, i.e., 
the frequency of the circulating currents for the electron vortex beams is zero. 
The cyclotron orbits of electrons with $L_z=0$ cuts the origin of the coordinate. 
In this case the change of the azimuthal angle of the electrons for one cyclotron orbit is $\pi$, 
hence the frequency of the circulating current of the electron vortex beam becomes 
half the cyclotron frequency $\omega_c/2$, i.e., Larmor frequency. 
In summary, the average circulating current for one electron in the classical electron vortex becomes 
\begin{subequations}
\begin{align}
I_c^>&=\frac{1}{2\pi} \omega_c \mbox{ for } L_z >0, \\
I_c^0&=\frac{1}{4\pi} \omega_c \mbox{ for } L_z =0, \\
I_c^<&=0  \mbox{ for } L_z <0.
\end{align}
\end{subequations}
This relation is equivalent to the quantum relation in Eq. (4) of Ref. \cite{SchttschneiderNC}. 
Thus the frequencies of the circulating current around the beam axis for the electron vortex beam 
is also well explained by the classical electron vortex. 

%We will consider the situation that electrons are moving in their cyclotron orbits with different centers and 
%many enough to make classical vortex-like wave similar situation that water molecules follows circular orbits to 
%cause a deep-water wave. 
%The three types of classical electron vortex waves (beams) are created according to the sign of 
%the canonical angular momentum of constituent electrons in the symmetric gauge. 

Now we will discuss the time dependence of the kinetic angular momentum of the classical electron vortex.  
It is surprising that the rotationally symmetric electron vortex beam has time-dependent kinetic angular momentum 
about its center contrast to the fact that the rotationally symmetric classical cyclotron motion 
has the constant kinetic angular momentum about its center.   
The expectation value of the kinetic angular momentum for the quantum mechanical vortex solution 
was shown to be time-dependent in general.  
The time-dependence of the quantum kinetic angular momentum is represented by \cite{Greenshields}
\begin{eqnarray}
\label{eq:TDKAM}
\langle L_z^{\mbox{\scriptsize{kin}}} \rangle(t) = \tilde{L}_z^{\mbox{\scriptsize{kin}}} + 
\left( \langle L_z^{\mbox{\scriptsize{kin}}} \rangle(0) - \tilde{L}_z^{\mbox{\scriptsize{kin}}} \right) \cos{(\omega_c t)},
\end{eqnarray}
where $\langle L_z^{\mbox{\scriptsize{kin}}} \rangle(t)$ is the expectation value of the kinetic angular momentum at 
time $t$ and $\tilde{L}_z^{\mbox{\scriptsize{kin}}}$ corresponds to the classical value of the kinetic angular momentum 
for an electron in a rotational 
motion about the origin of the coordinate with the radius $\tilde{\rho}=\sqrt{R^2+ R^2_{\mbox{\scriptsize{cen}}}}$. 
The kinetic angular momentum in Eq. (\ref{eq:TDKAM}) is constant for the special case of 
$\langle L_z^{\mbox{\scriptsize{kin}}} \rangle (0) = \tilde{L}_z^{\mbox{\scriptsize{kin}}}$, in which the vortex wave 
function is the Landau energy eigenstate (Landau state), however, it shows time-dependence for 
$\langle L_z^{\mbox{\scriptsize{kin}}} \rangle (0) \neq \tilde{L}_z^{\mbox{\scriptsize{kin}}}$, which describes 
a general electron vortex beams. 

%The apparent motion of the classical electron vortex, which is constructed by uniformly distributed electrons on their 
%cyclotron orbits, however, is a rotational motion around the center of the classical electron vortex at a constant rate. 
%This suggests that the average kinetic angular momentum of the classical electron vortex per one electron is a constant of motion.
The kinetic angular momentum of the classical electron vortex is defined by the average of the kinetic angular momenta 
of constituent electrons. 
The time-dependence of the kinetic angular momentum of each electron is given by the time behavior of the squared radius of 
the electron as Eqs. (\ref{eq:KAM}) and (\ref{eq:CRDE}).   
The squared radius of the classical electron vortex $\rho^2_V$ is defined by the average of the squared radii of constituent electrons 
$\rho^2$. The $\rho^2_V$ is equal to the average of $\rho^2$ over one cyclotron orbit because of the azimuthal symmetry of the vortex. 
The squared radius of the classical electron vortex is given by
\begin{eqnarray}
\rho^2_V=\langle \rho^2 \rangle_{c} = \langle ({\bf R}+{\bf R}_{\mbox{\scriptsize{cen}}} ) \cdot 
({\bf R}+{\bf R}_{\mbox{\scriptsize{cen}}} ) \rangle_c,
\end{eqnarray}
where ${\bf R}$ is the displacement of the electron from the center of the cyclotron orbit, 
${\bf R}_{\mbox{\scriptsize{cen}}}$ is the position vector of the center of the cyclotron orbit 
from the center of the classical vortex, and $\langle \cdot \rangle_{c}$ means the average over one cyclotron orbit.  
For our classical electron vortex with uniform distributions of electrons on each cyclotron orbit, 
$\langle {\bf R}\rangle_c=0$, which implies that $\langle \rho \rangle_{c}$ is equal to $\tilde{\rho}$ 
and the average kinetic angular momentum $\langle L_z^{\mbox{\scriptsize{kin}}} \rangle_c$ is the same as $\tilde{L}_z^{\mbox{\scriptsize{kin}}}$ so that 
the kinetic angular momentum remains constant. 

\begin{figure}[htbp]
\subfloat[]{
\includegraphics[height=1.0in]{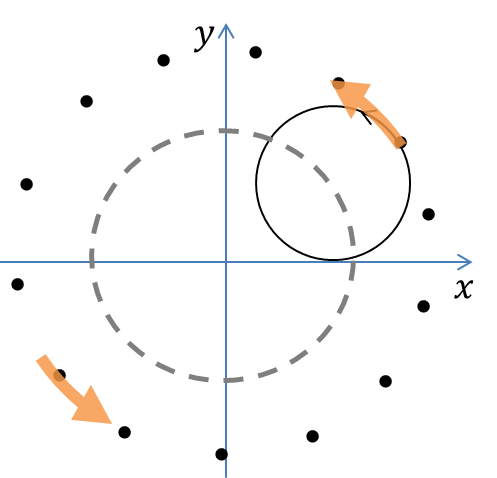}}
%\caption{The cyclotron orbit for $L_z<0$}
\subfloat[]{
\includegraphics[height=1.0in]{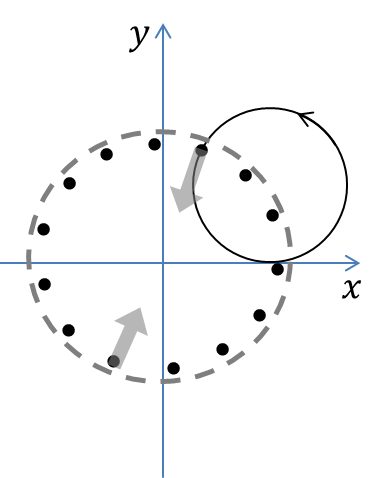}}
\subfloat[]{
\includegraphics[height=1.0in]{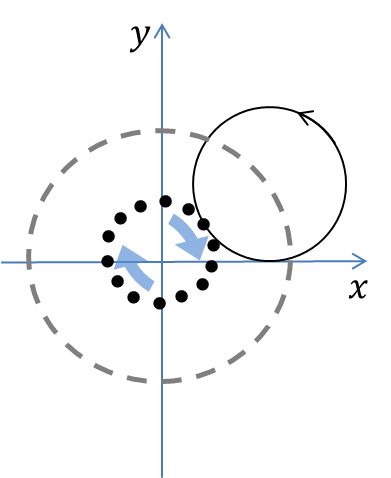}}

\caption{(Color online) Simple example to show the oscillating behavior of the radius of the classical electron vortex 
for $L_z<0$. Here (a) and (c) show the azimuthal currents at outmost and inmost positions, respectively, and 
(b) shows the instant, at which the azimuthal current is zero.}
\label{fig:VEX}
\end{figure}

Then the question is how one can create a classical electron vortex with time-dependent kinetic angular momentum. 
That classical electron vortex is created in the case that 
the distribution of the constituent electrons on each cyclotron orbit is not uniform 
and satisfies $\langle {\bf R}\rangle_c \neq 0$. 
One simple example is the case that there is only one electron on each cyclotron orbit as shown in Fig. \ref{fig:VEX}. 
We suppose that each electron is initially on the outmost point of each cyclotron orbit as in Fig. \ref{fig:VEX} (a).   
Then it is obvious that the radius of the classical electron vortex created by the cyclotron motions of these 
electrons becomes time-dependent as the constituent electrons rotate in their cyclotron orbits as shown in Fig. 
\ref{fig:VEX} (b) and (c). 
This makes the diamagnetic angular momentum time-dependent and as a result, the kinetic angular momentum 
becomes time-dependent. 
The time-dependence of the squared radial distance of one electron is governed by Eq. (\ref{eq:CRDE}), 
which is equivalent to Eq. (11) of Ref. \cite{Greenshields}, because the vortex motion is created 
by the motions of the constituent electrons.

As another feature of the classical electron vortex, let us study the relation between the kinetic angular momentum 
and the energy of the electron vortex. 
It is expected that that relation of the electron vortex per 
one electron is equal to that of one cyclotron orbit because the electron vortex is composed of the cyclotron 
motion of the constituent electrons.
 
In one electron case, the kinetic angular momentum of an electron in its cyclotron motion becomes a constant of motion 
for the coordinate whose origin is the center of the motion, i.e., $x_0=y_0=0$ in Eq. (\ref{eq:2DSOL}). 
Then the diamagnetic angular momentum 
is the same as the canonical angular momentum such that the kinetic angular momentum is 
$2 L_z= m \omega_c R^2$. 
Therefore, the rotational kinetic energy $E$ of the electron is determined by the kinetic angular momentum as
\begin{eqnarray}
\label{eq:CEKAM}
E=\frac{|e|B}{2m }L_z^{\mbox{\scriptsize{kin}}}=\frac{1}{2}\omega_c L_z^{\mbox{\scriptsize{kin}}}.
\end{eqnarray}

In the classical electron vortex, even though the centers of the constituent cyclotron motions constructing 
the classical electron vortex are not 
the origin of the coordinate, the kinetic angular momentum can remain constant as we have seen for the 
classical electron vortex with $\langle {\bf R} \rangle_c=0$. 
In this classical electron vortex the average kinetic angular momentum of one 
electron, $\langle L_z^{\mbox{\scriptsize{kin}}} \rangle_e$, is calculated as 
\begin{eqnarray}
\langle L_z^{\mbox{\scriptsize{kin}}} \rangle_e= \langle {\bf r}\times m {\bf v} \rangle_e 
=\langle {\bf R}\times m{\bf v} \rangle_e=mR^2 \omega_c,
\end{eqnarray}  
using the uniform distribution of electrons in one cyclotron orbit, 
because $\langle m{\bf v} \rangle_e$ is zero, where $\langle \cdot \rangle_e$ means the classical average 
of $\cdot$ per one electron.  
Hence the energy per electron $E_e$ is given by the following 
relation 
\begin{eqnarray}
\label{eq:CVEAM}
E_e=\frac{1}{2}\omega_c \langle L_z^{\mbox{\scriptsize{kin}}} \rangle_e.
\end{eqnarray}

For the Landau state, i.e., $\langle L_z^{\mbox{\scriptsize{kin}}} \rangle (0) =\tilde{L}_z^{\mbox{\scriptsize{kin}}}$, 
the kinetic angular momentum of the electron vortex beams is also time-independent and the energy eigenvalue is 
represented by the expectation value of the kinetic angular momentum as \cite{Li}
\begin{eqnarray}
\label{eq:QVEAM}
E_{n,l}=\left( n + \frac{|l|}{2}+\frac{m}{2} +\frac{1}{2}\right) \hbar \omega_c
=\frac{1}{2}\omega_c \langle L_z^{\mbox{\scriptsize{kin}}} \rangle(0)
\end{eqnarray}
 This relation is equivalent to the relation in Eq. (\ref{eq:CVEAM}) and shows that the energy and kinetic angular momentum 
relations are equivalent in classical and quantum electron vortices.

For the generic time-dependent case, the time-average of the kinetic angular momentum must be used, which becomes 
the kinetic angular momentum of the cyclotron motion with the center at the origin of the coordinate 
for both quantum and classical electron vortices. 
This is because the time average of the cyclotron motion of one electron is equal to the average over the uniformly 
distributed electrons in their cyclotron motions. 
Hence the energy and kinetic angular momentum relation in Eqs. (\ref{eq:CVEAM}) and (\ref{eq:QVEAM}) are 
satisfied for all electron vortices.

The rotational motion of the classical electron vortex naturally leads to the parallel axis theorem with the 
observation that the moment of inertia $I$ about the propagating direction of the classical electron vortex ($z$-axis) is 
\begin{eqnarray}
\label{eq:PAT}
I= m \rho^2=m(R^2+R_{\mbox{\scriptsize{cen}}}^2)
\end{eqnarray}
For the uniform distribution of the constituent electrons, the $R_{\mbox{\scriptsize{cen}}}$ becomes the center of 
mass of the electrons on each cyclotron orbit. 
In this case, the equality in Eq. (\ref{eq:PAT}) represents the usual parallel axis theorem, in which 
the moment of inertia about any axis is the sum of the moment of inertia of one particle with the total mass of the 
system about the same axis and the moment of inertia about the parallel axis through the center of mass. 
When the distribution of the constituent electrons are not uniform, the center of mass is not $R_{\mbox{\scriptsize{cen}}}$
and time-dependent, so that the parallel axis theorem becomes 
\begin{eqnarray}
I=m \rho^2= m(\langle \tilde{R} \rangle^2 + R_{\mbox{\scriptsize{C.M.}}}^2), 
\end{eqnarray}
where $R_{\mbox{\scriptsize{C.M.}}}$ is the radial distance from the origin of the coordinate to the 
center of mass and $\langle \tilde{R}\rangle^2 = \int \tilde{\bf R}\cdot \tilde{\bf R}dm/ m$, 
where $\tilde{\bf R}$ is the radial vector from the center of mass to the position of the infinitesimal mass $dm$. 
The $R_{\mbox{\scriptsize{C.M.}}}$ and $\tilde{\bf R}$ can be definitely time-dependent and so is $I$. 
This parallel axis theorem shows the same feature in Ref. \cite{GreenshieldsNJ}.

\section{Conclusion}
We have proposed the classical electron vortex model for the electron vortex beams in a uniform magnetic field. 
The classical electron vortex is constructed by the collective motion of the constituent electrons 
in their cyclotron motions under the uniform magnetic field with constant canonical angular momentum in the symmetric gauge. 
The basic features of the electron vortex beam were understandable by the classical electron vortex. 
The canonical angular momentum of one electron in a uniform magnetic field has three categories, 
positive, zero, and negative, according to the distance of the center of the cyclotron orbit from 
the center of the classical electron vortex. 
Then it is shown that the three kinds of the classical electron vortex, 
which is qualitatively equivalent to quantum ones, exist according to the three categories of the canonical 
angular momentum. 
The energy of the classical electron vortex, per one constituent electron, is shown to be the energy of the classical 
cyclotron motion of one electron as expected in the classical physics, independent on the value of 
the canonical angular momentum. 

The surprising time-dependence of the kinetic angular momentum of the electron vortex beams was explained by the mismatch between 
the centers of the cyclotron orbits and the classical electron vortex for the distribution of electrons in which 
the average of the displacements of the electrons from the center of the cyclotron orbit is not the 
center of the cyclotron orbit. This mismatch also naturally induced the parallel axis theorem of the kinetic 
angular momenta of the electron vortex beams. 
Additionally, the three categories of the angular frequencies of the azimuthal currents of the electron vortex beams 
are also explained in the classical electron vortex by using the corresponding circulating 
currents of one electron.

%The classical dynamics seems probable not to explain these three frequencies because the classical electron
%in a uniform magnetic field rotates in a circular orbit only with the cyclotron frequency.  

The results in this paper suggest that the physics of the electron vortex can be 
understood by the classical motion of electrons. 
We hope that the classical electron vortex model would help to understand the abundant new physics 
intuitively and to differentiate classical and quantum physics in electron vortex beams.  

 \section*{Acknowledgements}

T. Choi was supported by a research grant from Seoul Women's University(2015) and 
Y. D. Han was supported by a research grant from Woosuk University.


\begin{thebibliography}{99}

\bibitem{Bliokh07} K. Y. Bliokh {\it et al.}, Phys. Rev. Lett. {\bf 99}, 190404 (2007). 
\bibitem{Uchida} M. Uchida and A. Tonomura, Nature (London) {\bf 464}, 737 (2010).
\bibitem{McMorrain} B. J. McMorran, A. Agrawal, I. M. Anderson, A. A. Herzing, H. J. Lezec, 
J. J. McClelland, and J. Unguris, Science {\bf 331}, 192 (2011).
\bibitem{Verbeeck10} J. Verbeeck, H. Tian, and P.Schattschneider, Nature (London) {\bf 467}, 301 (2010). 
\bibitem{VerbeeckUl} P. Schattschneider and J. Verbeeck, Ultramicroscopy {\bf 111}, 1461 (2011).
\bibitem{Verbeeck11} J. Verbeeck, P. Schattschneider, S. Lazar, M. StogerP\"llach, S. L\"offler, 
A. Steiger-Thirsfeld, and G. Van Tendeloo, Appl. Phys. Lett. {\bf 99}, 203109 (2011). 
\bibitem{Bliokh11} K. Y. Bliokh, M. R. Dennis, and F. Nori, Phys. Rev. Lett. {\bf 107}, 174802 (2011). 
\bibitem{Guzzinati} G. Guzzinati, P. Schattachneider, K. Y. Bliokh, F. Nori, and J. Verbeeck, Phys. Rev. Lett. 
{\bf 110}, 093601 (2013). 
\bibitem{Lloyd} S. Lloyd, M. Babiker, and J. Yuan, Phys. Rev.Lett. {\bf 108}, 044801 (2012).
\bibitem{Karimi} E. Karimi, L. Marrucci, V. Grillo, and E. Santamato, Phys. Rev. Lett. {\bf 108}, 074802 (2012).
\bibitem{Gallatin12} G. M. Gallatin and B. McMorran, Phys. Rev. A {\bf 86}, 012701 (2012). 
\bibitem{SchttschneiderNC} P. Schttschneider, Th. Schachinger, M. St\"oger-Pollach, S. L\"offler, 
A. Steiger-Thirsfeld, K. Y. Bliokh, and F. Nori, Nat. Comm. {\bf 5}, 4586 (2014). 
\bibitem{Ballentine} L. E. Ballentine, {\it Quantum Mechanics : A Modern Development} World Scientific 
Co. Pte. Ltd. Singapore 1988. 
\bibitem{Greenshields} C. R. Greenshields, R. L. Stamps, S. Franke-Arnold, and S. M. Barnett, 
Phys. Rev. Lett. {\bf 113}, 240404 (2014).
\bibitem{Greenshields15} C. R. Greenshields, S. Franke-Arnold, and R. L. Stamps, New J. Phys. {\bf 17}, 
093015 (2015).
\bibitem{GreenshieldsNJ} C. R. Greenshields, R. L. Stamps, and S. Franke-Arnold, New J. Phys. 
{\bf 14}, 103040 (2012). 
\bibitem{BliokhX} K. Y. Bliokh, P. Schattschneider, J. Verbeeck, and F. Nori, 
Phys. Rev. X {\bf 2}, 041011 (2012).
\bibitem{Li}C-F. Li and Q. Wang, Physica B {\bf 269}, 22 (1999). 


  
 %\bibitem{Coecke} B. Coecke and R. Lal, Phys. Rev. Lett. {\bf 108}, 200403 (2012).
 %\bibitem{Simon} C. Simon, V. Bu$\breve{z}$ek, and N. Gisin, Phys. Rev. Lett. {\bf 87},
% 170405 (2001); {\bf 90}, 208902 (2003); P. B$\acute{o}$na, {\it ibid.}.

\end{thebibliography}
\end{document}